\documentclass[aps,prl,twocolumn,floats,epsfig,showpacs]{revtex4}
\usepackage{bbm}
\usepackage{ifpdf}
\usepackage{graphicx,epsfig}

\begin{document} \draft 
\title{Nested Cluster Algorithm for Frustrated Quantum Antiferromagnets}
\author{M.\ Nyfeler, F.-J.\ Jiang, F.\ K\"ampfer, and U.-J.\ Wiese}
\address{Institute for Theoretical Physics, Bern University,
Sidlerstrasse 5, 3012 Bern, Switzerland}

\begin{abstract}

Simulations of frustrated quantum antiferromagnets suffer from a severe sign 
problem. We solve the ergodicity problem of the loop-cluster algorithm in a 
natural way and apply a powerful strategy to address the sign problem. For the 
spin $\frac{1}{2}$ Heisenberg antiferromagnet on a kagom\'e and on a frustrated
square lattice, a nested cluster algorithm eliminates the sign problem for 
large systems. The method is applicable to general lattice geometries but 
limited to moderate temperatures.

\end{abstract} 

\pacs{75.10Jm, 75.40Mg, 75.50.Ee}

\maketitle

Monte Carlo calculations are a powerful tool for first principles 
investigations of strongly coupled quantum systems. Early simulations of 
quantum spin systems were performed with local Metropolis-type algorithms 
\cite{Reg88}. They suffered from critical slowing down and were thus limited to
rather high temperatures. Cluster algorithms perform non-local updates and are 
capable of substantially reducing critical slowing down. Such algorithms 
were first developed by Swendsen and Wang for discrete classical Ising and 
Potts spins \cite{Swe87} and then generalized by Wolff \cite{Wol89} to 
classical spins with a continuous $O(N)$ symmetry. Improved estimators which
average over an exponentially large number of configurations at polynomial cost
are an additional benefit of cluster algorithms. The first cluster algorithm
for the spin $\frac{1}{2}$ quantum Heisenberg model was developed in 
\cite{Wie92}. While that algorithm works efficiently only for quantum spin
chains, the loop-cluster algorithm \cite{Eve93} is efficient also in higher 
dimensions, and was first applied to the 2-d spin $\frac{1}{2}$ Heisenberg 
antiferromagnet on a square lattice in \cite{Wie94}. The continuous-time 
variant of the algorithm eliminates the Suzuki-Trotter time-discretization 
error and can reach very low temperatures \cite{Bea96}. This method has also 
been used to simulate systems on very large lattices \cite{Kim98} and with 
very long correlation lengths \cite{Bea98}. An elegant and powerful related 
method based on stochastic series expansion is available as well \cite{San97}.

Unfortunately, in many cases of physical interest, including frustrated quantum
spin systems, quantum Monte Carlo calculations suffer from a very severe sign 
problem. Using an improved estimator, the sign problem of the 2-d classical 
$O(3)$ model at vacuum angle $\theta = \pi$ has been addressed with a variant 
of the Wolff cluster algorithm \cite{Bie95}. In that case, some clusters are 
half-instantons also known as merons. Flipping a meron-cluster leads to a 
sign-change of the Boltzmann weight and hence to an exact cancellation between 
two configurations. As a consequence, only configurations without 
meron-clusters contribute to the partition function. Restricting the simulation
to those configurations eliminates the sign problem, since all configurations
in the zero-meron sector have a positive sign. The meron concept has been 
generalized to fermionic systems \cite{Cha99} and the meron-cluster algorithm 
has been used to solve a number of very severe fermion sign problems 
\cite{Cha00,Osb02,Cha03}. Unfortunately, the meron-cluster algorithm is not 
generally applicable. In fact, as shown in \cite{Tro05}, some sign problems are
NP-complete. Hence, a hypothetical method that can solve any sign problem would
solve all NP-complete problems in polynomial time. This would imply the
equality of the complexity classes NP$=$P. Since it is generally believed that
NP$\neq$P, it is expected that a universally applicable method that solves all
sign problems cannot exist. In this paper we construct a nested cluster
algorithm which, for the first time, is capable of eliminating severe sign
problems of frustrated antiferromagnets at least at moderate temperatures.

Let us consider the antiferromagnetic spin $\frac{1}{2}$ quantum Heisenberg 
model with the Hamiltonian
\begin{equation}
H = \sum_{x, y \in \Lambda} J_{xy} \vec S_x \cdot \vec S_y.
\end{equation}
Here $\vec S_x$ is a quantum spin operator located at the site $x$ of a lattice
$\Lambda$, and  $J_{xy} > 0$ is the antiferromagnetic exchange coupling between
a pair of spins located at the sites $x$ and $y$. Although our method can be 
applied directly in the Euclidean time continuum, in order to ease its 
implementation we work in discrete time. Depending on the lattice geometry, the
Hamiltonian $H = H_1 + H_2 + ... + H_M$ is expressed as a sum of $M$ terms 
$H_i$ which leads to a Suzuki-Trotter decomposition of the partition function
\begin{eqnarray}
\label{Trotter}
Z\!\!\!&=&\!\!\!\mbox{Tr} \exp(- \beta H) \nonumber \\
\!\!\!&=&\!\!\!\lim_{\varepsilon \rightarrow 0}
\mbox{Tr} \left[\exp(- \varepsilon H_1) \exp(- \varepsilon H_2) ...
\exp(- \varepsilon H_M)\right]^N\!\!.
\end{eqnarray}
Here the inverse temperature $\beta = 1/T = N \varepsilon$ represents the 
extent of a periodic Euclidean time interval, which is divided into $N$ 
discrete time steps of size $\varepsilon$. Each $H_i$ is a sum of mutually 
commuting pair interactions $h_{xy} = J_{xy} \vec S_x \cdot \vec S_y$ on a set 
of disconnected bonds. Inserting complete sets of spin states 
$s_{x,t} = \pm \frac{1}{2} = \uparrow, \downarrow$ between the factors 
$\exp(- \varepsilon H_i)$ in eq.(\ref{Trotter}), the partition function is 
expressed as a path integral over all spin configurations $[s]$ \cite{Wie94}
\begin{equation}
Z = \sum_{[s]} \mbox{Sign}[s] \exp(- S[s]).
\end{equation}
The weight of a configuration is a product of contributions from individual 
space-time plaquettes corresponding to the two-spin transfer matrix elements
$\langle s_{x,t} s_{y,t}|\exp(- \varepsilon h_{xy})|s_{x,t+1} s_{y,t+1} \rangle$.
In the basis $|\uparrow \uparrow \rangle$, $|\uparrow \downarrow \rangle$,
$|\downarrow \uparrow \rangle$, $|\downarrow \downarrow \rangle$, up to an
irrelevant overall factor, the two-spin transfer matrix takes the form 
\begin{equation}
\label{weights}
\exp(- \varepsilon h_{xy}) = \left(\begin{array}{cccc} A & 0 & 0 & 0 \\
0 & A + B & - B & 0 \\ 0 & - B & A + B & 0 \\ 0 & 0 & 0 & A \end{array}\right),
\end{equation}
with $A = \exp(- \varepsilon J_{xy}/2)$ and $B = \sinh(\varepsilon J_{xy}/2)$.
The off-diagonal transfer matrix elements are negative. The product of the 
negative signs over all space-time plaquettes defines the total
$\mbox{Sign}[s] = \pm 1$ of a spin configuration. The remaining factor 
$\exp(- S[s])$ represents a positive Boltzmann weight which can be interpreted 
as a probability and thus can be used for importance-sampling in a Monte 
Carlo simulation. 

When one samples the system using the positive weight $\exp(- S[s])$, one must 
include $\mbox{Sign}[s]$ in the measured observables $O[s]$ and expectation 
values are given by
\begin{equation}
\langle O \rangle = \frac{1}{Z} \sum_{[s]} O[s] \ \mbox{Sign}[s] \exp(- S[s]) =
\frac{\langle O \ \mbox{Sign} \rangle_+}{\langle \mbox{Sign} \rangle_+}.
\end{equation}
Here the index $+$ refers to expectation values in the simulated ensemble with 
positive Boltzmann weights and partition function 
$Z_+ = \sum_{[s]} \exp(- S[s])$ such that
\begin{eqnarray}
\label{sign}
\langle \mbox{Sign} \rangle_+&=& 
\frac{1}{Z_+} \sum_{[s]} \mbox{Sign}[s] \exp(- S[s]) = \frac{Z}{Z_+} 
\nonumber \\
&\sim&\exp(- \Delta f \beta V).
\end{eqnarray}
Here $V$ is the spatial volume and $\Delta f$ is the difference between the 
free energy densities of the original ensemble with the weight 
$\mbox{Sign}[s] \exp(- S[s])$ and the simulated ensemble with the positive
weight $\exp(- S[s])$. The expectation value of the sign is exponentially small
in the space-time volume $\beta V$. Since it is obtained as a Monte Carlo 
average of contributions $\mbox{Sign}[s] = \pm 1$, one needs an exponentially 
large statistics in order to accurately measure 
$\langle \mbox{Sign} \rangle_+$. This is impossible in practice and gives rise 
to a very severe sign problem.

How can one increase the statistics by an exponential factor without investing 
more than a polynomial numerical effort? The meron-cluster algorithm
\cite{Bie95,Cha99} achieves this by constructing an improved estimator for the 
sign. Like the meron-cluster algorithm, the method presented here is based on 
the loop-cluster algorithm \cite{Eve93} which decorates a spin configuration 
with bonds connecting spins to closed loop-clusters. The four spins on a 
space-time plaquette are connected in pairs. In fact, $A$ and $B$ in 
eq.(\ref{weights}) represent weights of two possible bond configurations on a 
space-time plaquette. The weight $A$ corresponds to bonds connecting the spins 
$s_{x,t}$ and $s_{y,t}$ with their time-like neighbors $s_{x,t+1}$ and 
$s_{y,t+1}$, while $B$ corresponds to space-like bonds connecting 
$s_{x,t}$ with $s_{y,t}$ and $s_{x,t+1}$ with $s_{y,t+1}$. Sites connected by 
bonds form a closed oriented loop-cluster. Up to an overall spin-flip of the 
entire cluster, the spin configuration on a cluster is determined by the 
cluster geometry. Time-like bonds connect parallel spins, while space-like 
bonds connect anti-parallel spins. Integrating out the spins, the partition
function can be expressed as a sum over bond configurations $[b]$
\begin{equation}
\label{bondZ}
Z = \sum_{[b]} \mbox{Sign}[b] A^{n_A} B^{n_B} 2^{N_{\cal C}}.
\end{equation}
Here $n_A$ is the number of time-like and $n_B$ is the number of space-like 
plaquette break-ups, while $N_{\cal C}$ is the number of loop-clusters. The
factor $2^{N_{\cal C}}$ arises because each cluster has two possible spin
orientations. The partition function can be sampled by a Metropolis 
update of the plaquette break-ups. Remarkably, while the original cluster 
algorithm which operates on spins and bonds never changes the sign and is thus 
not ergodic \cite{Hen00}, the algorithm which operates only on bonds (after the
spins have been integrated out) is ergodic and still avoids unnatural freezing.
Interestingly, $\mbox{Sign}[s]$ remains invariant under cluster flips, i.e.\ 
all clusters are non-merons. However, in this case the meron-cluster algorithm 
does not solve the sign problem because almost half of the configurations in 
the zero-meron sector have a negative sign \cite{Hen00}. Since it does not 
change under spin flips, $\mbox{Sign}[s] = \mbox{Sign}[b]$ is uniquely 
determined by the bond configuration. It is important to note that the sign can
be expressed as a product of cluster signs $\mbox{Sign}[b] = \prod_{\cal C} 
\mbox{Sign}_{\cal C}$. Depending on the orientation of a cluster, each 
space-like break-up contributes a factor $\pm i$ to the two clusters traversing
the corresponding space-time plaquette. By construction, each cluster traverses
an even number of space-like break-ups, and hence 
$\mbox{Sign}_{\cal C} = \pm 1$.

We distinguish space-time plaquettes shared by two different clusters from 
internal plaquettes belonging entirely to one cluster. Updating the break-up on
a space-time plaquette shared by two different clusters does not lead to a 
sign-change. Only updates of cluster-internal plaquettes may change the sign. 
We apply the following method to construct an improved estimator for the 
sign. Once a statistically independent bond configuration has been produced 
by the cluster algorithm, we perform an inner Monte Carlo simulation by
updating only the cluster-internal plaquette break-ups. Each cluster ${\cal C}$
defines the set of lattice sites $\Lambda_{\cal C}$ contained in ${\cal C}$. The
inner Monte Carlo algorithm generates clusters with different orientations that
visit all sites of $\Lambda_{\cal C}$ in different orders, thus contributing 
different values of $\mbox{Sign}_{\cal C}$. In this process, break-ups that 
lead to the decomposition of $\Lambda_{\cal C}$ into separate clusters must be 
rejected. The inner Monte Carlo algorithm estimates an average
$\langle \mbox{Sign}_{\cal C} \rangle_i$ for each set of sites
$\Lambda_{\cal C}$. Since the different sets are independent, the improved 
estimator of the sign is given by
\begin{equation}
\langle \mbox{Sign} \rangle_i = \prod_{\Lambda_{\cal C}} 
\langle \mbox{Sign}_{\cal C} \rangle_i.
\end{equation}
Remarkably, the nesting of an outer and an inner cluster algorithm achieves 
exponential error reduction at polynomial cost. A similar strategy was very 
successfully applied to the measurement of exponentially suppressed Wilson 
loops in lattice gauge theory \cite{Lue01} as well as to quantum impurity
models \cite{Yoo05}. Correlation functions 
and susceptibilities can also be measured with improved estimators. Let us
consider the staggered magnetization operator $\vec M_s = \sum_x z_x \vec S_x$. 
Here $z_x$ is a stagger factor depending on the sub-lattice to which the site 
$x$ belongs. The corresponding staggered susceptibility 
\begin{equation}
\chi_s = \frac{\langle M_s^2 \mbox{Sign} \rangle_+}
{\beta V \langle \mbox{Sign} \rangle_+} =
\frac{\langle \langle M_s^2 \mbox{Sign} \rangle_i \rangle_+}
{\beta V \langle \langle \mbox{Sign} \rangle_i \rangle_+}.
\end{equation}
is obtained from an improved estimator which is given in terms of 
$M_s = \sum_{\cal C} M_{s{\cal C}}$ with 
$M_{s{\cal C}} = \sum_{(x,t) \in {\cal C}} z_x s_{x,t}$ as
\begin{equation}
\langle M_s^2 \mbox{Sign} \rangle_i = \sum_{\Lambda_{\cal C}}
\langle M_{s{\cal C}}^2 \mbox{Sign}_{\cal C} \rangle_i 
\prod_{\Lambda_{\cal C'} \neq \Lambda_{\cal C}} \langle \mbox{Sign}_{\cal C'} \rangle_i.
\end{equation}

In which cases will the nested cluster algorithm eliminate or at least 
substantially reduce the sign problem? Since some sign problems are NP-hard, it
is expected that any method will fail at least in those cases. The nested 
cluster algorithm fails to solve the sign problem when a cluster fills almost 
the entire volume, because then the inner Monte Carlo algorithm becomes 
inefficient. Since large clusters necessarily arise in the presence of large 
correlation lengths, the nested cluster algorithm does not work efficiently in 
low-temperature ordered phases. 

Even in the absence of long-range order, cluster algorithms may
become inefficient if the clusters grow to unphysically large sizes beyond the 
physical correlation length. This potential problem is prevented when there is
a reference configuration that limits cluster growth \cite{Cha03}. For the 
antiferromagnet on the square lattice the reference configuration is given by
the classical N\'eel state, i.e.\ all spins in a loop-cluster are in a 
staggered pattern. The cluster-size squared is then tied to the staggered 
susceptibility which protects the clusters from growing to unphysically large 
sizes. Also for frustrated systems it is natural to consider a classical ground
state as a reference configuration. When one quantizes the spins along a local 
quantization axis in the direction of the spin orientation in the classical 
ground state, an interesting algorithm with open string-clusters emerges. The 
spins in each cluster are in the reference configuration and hence these 
clusters are protected from becoming unphysically large. However, the 
meron-concept does not apply to the open string-clusters, i.e.\ when these 
clusters are flipped, they are not independent but affect each other in their 
effect on the sign. Remarkably, one can still integrate out the spins
analytically. This glues the open string-clusters together to the closed 
loop-clusters of the algorithm discussed before. While typical closed 
loop-clusters are hence larger than the correlation length corresponding to the
classical order, they still represent physical correlated regions. In fact, 
they grow up to the length scale at which the signs, which are a manifestation 
of quantum entanglement, decorrelate.

Even if the typical cluster-size is moderate, the inner Monte Carlo algorithm
may not lead to an efficient cancellation of signs. For example, there are 
cases in which the improved estimator $\langle \mbox{Sign} \rangle_i$ is not 
positive. Still, if such cases are rare, the sign problem is substantially
reduced. In order to optimize the performance of the algorithm, the numerical
effort invested in the inner and outer Monte Carlo procedures must be properly 
balanced against each other. It pays off to invest a larger number of inner 
Monte Carlo sweeps on the larger sets $\Lambda_{\cal C}$. In any case, the 
efficiency of the nested cluster algorithm must be investigated on a case by 
case basis.

We now consider the Heisenberg antiferromagnet with uniform nearest-neighbor 
coupling $J_{xy} = J$ on the lattices illustrated in figure 1. The frustrated 
square lattice has an additional coupling $J'$ along the diagonals. 
\begin{figure}
\includegraphics[width=0.43\textwidth]{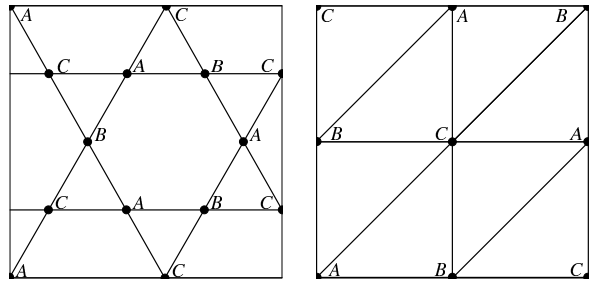}
\caption{\it kagom\'e lattice (left) and frustrated square (or anisotropic
triangular) lattice (right) consisting of three sub-lattices $A, B, C$.}
\end{figure}
We have simulated large kagom\'e lattices with up to $V \approx 1000$ spins at 
moderate temperatures with $\beta J \approx 1$. Figure 2 shows the probability 
distribution of the improved estimator $\langle \mbox{Sign} \rangle_i$. 
Although sometimes it is negative, it still leads to an accurate determination 
of the average sign. 
\begin{figure}
\includegraphics[width=0.42\textwidth]{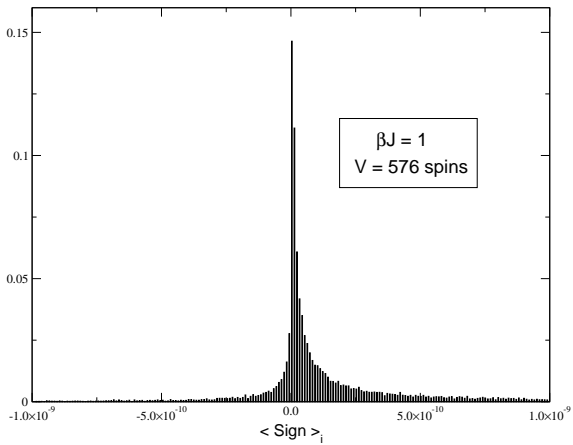}
\caption{\it Probability distribution of $\langle \mbox{Sign} \rangle_i$ for 
the kagom\'e lattice with $V = 576$ spins and $\beta J = 1$.}
\end{figure}
We consider $M_s$ with $z_x = 1, -1, 0$ on sub-lattice $A, B, C$, respectively,
which may signal coplanar spin order. As shown in figure 3, 
with increasing volume $V$ both $\langle \mbox{Sign} \rangle_+$ and 
$\langle M_s^2 \mbox{Sign} \rangle_+$ decrease dramatically over numerous 
orders of magnitude, but are still accurately accounted for by the nested 
cluster algorithm. For example, with $V = 882$ spins 
$\langle \mbox{Sign} \rangle_+ = 2.09(8) \times 10^{-14}$. A brute force 
approach would require an astronomical statistics of about $10^{30}$ sweeps in 
order to achieve a similar precision.
\begin{figure}
\includegraphics[width=0.42\textwidth]{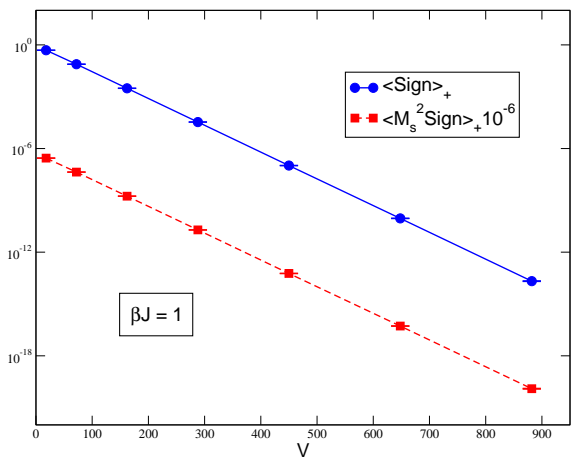}
\caption{\it Volume-dependence of $\langle \mbox{Sign} \rangle_+$ and 
$\langle M_s^2 \mbox{Sign} \rangle_+$ (rescaled by $10^{-6}$) for the kagom\'e 
lattice with $\beta J = 1$.}
\end{figure}
Figure 4 shows the coplanar staggered susceptibility $\chi_s$ compared to the 
collinear N\'eel susceptibility $\chi_N$. On the square lattice, 
frustration reduces the N\'eel order, while (at least for $J' = J/4$) the 
coplanar order is as weak as on the kagom\'e lattice (and practically 
indistinguishable from it in figure 4).
\begin{figure}[b]
\includegraphics[width=0.42\textwidth]{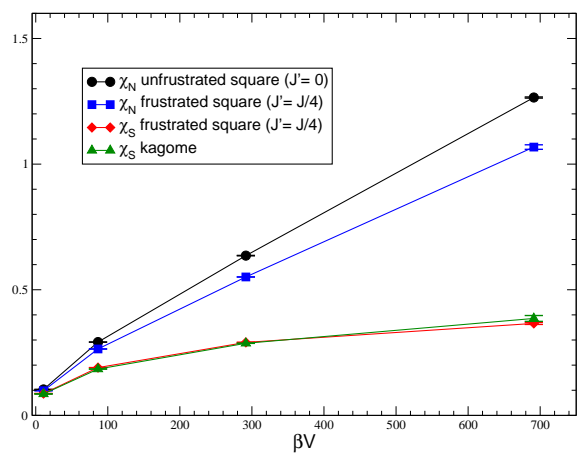}
\caption{\it Coplanar staggered susceptibility $\chi_s$ and collinear N\'eel 
susceptibility $\chi_N$ as functions of the space-time volume $\beta V$ for the
kagom\'e as well as the frustrated ($J' = J/4$) and unfrustrated ($J' = 0$) 
square lattice at fixed space/time aspect ratio $\sqrt{V}/\beta J = 20$.}
\end{figure}

To conclude, in contrast to other Monte Carlo methods, the nested cluster 
algorithm is capable of eliminating very severe sign problems for large 
systems, at least at moderate temperatures. This is useful, for example, for 
determining the couplings of frustrated magnets by comparison with experimental
finite temperature data. As we have demonstrated, although the nested cluster
algorithm cannot reach very low temperatures, by studying appropriate 
susceptibilities one may still obtain valuable insights concerning possible 
types of order. Applications to frustrated antiferromagnets on various lattice
geometries are currently in progress.

U.-J.\ W.\ likes to thank S.\ Chandrasekharan for a long-lasting fruitful and 
very inspiring collaboration on the sign problem. We also have benefited from 
interesting discussions with M.\ Troyer. This work was supported by the 
Schweizerischer Nationalfonds.

\end{document}